\documentclass[apj]{emulateapj}
\usepackage{graphics}

\shorttitle{Motion of BLR clouds}
\shortauthors{F. Khajenabi}

\begin{document}

\title{Effect of the drag force on the orbital motion of  the broad-line region clouds}

\author{Fazeleh Khajenabi }
\affil{Department of Physics, Faculty of Sciences, Golestan University, Gorgan 49138-15739, Iran \\ f.khajenabi@gu.ac.ir}

\begin{abstract}
We investigate orbital motion of cold clouds in the broad line region of active galactic nuclei subject to the gravity of a black hole and a force due to a nonisotropic central source and a drag force proportional to the velocity square. The intercloud is described using the standard solutions for the advection-dominated accretion flows. Orbit of a cloud decays because of the drag force, but the typical time scale of falling of clouds onto the central black hole is shorter comparing to the linear drag case.  This time scale is calculated when a cloud is moving through a static or rotating intercloud. We show that when the drag force is a quadratic function of the velocity, irrespective of the initial conditions and other input parameters, clouds will generally fall onto the central region much faster than the age of whole system and since cold clouds present in most of the broad line regions, we suggest that mechanisms for continuous creation of the clouds must operate in these systems. 
\end{abstract}


\section{Introduction}
%
%


Emission of broad-line region (BLR) of active galactic nuclei (AGNs) is explained by models which propose  continuous steady flows or presence a very large number of clouds which exhibit some bulk motion \citep[e.g.,][]{oster,netzerbook,Rees,Rees89}. In the early models for the BLRs, the clouds are moving inward or outward through a static or slowly moving background gas \citep[e.g.,][]{mathews74,mat75,mat79}. Subsequent studies showed that the clouds may exhibit orbital motion under the strong gravitational potential of a central object. \cite{kwan82} constructed a kinematic model in which the BLR clouds orbit the central object in nearly parabolic orbits and this model has been extended by \cite{carroll85} to include the effects of a finite infalling cloud number and size. There are considerable uncertainties about formation of the BLR clouds and their dynamical stability \citep[e.g.,][]{mat86,mat89}, however, many authors have successfully produced emission of BLR systems based on the discrete cloud concept \citep[e.g.,][]{cap79,cap80,cap81,frome,netzer2010}.

  Most of these models assume that the BLR clouds are pressure-confined, though a few authors argue that the clouds are transient rather than stable long-lived objects. Orbital motion of the BLR clouds is a rich source of information to estimate the mass of the central black hole. While early models  are considering gravitational force of the central black hole as a dominant force, it has been argued that BLR clouds are also subject to a force due to the intense radiation of a central source (e.g., an accretion disc) and the mass of the central black hole is underestimated if radiation pressure is neglected \citep[e.g.,][]{marconi,namekata}. Moreover, it has  been suggested  that the central radiation is nonisotropic  \citep{Liu11} and the orbits of BLR clouds are significantly modified when this feature of radiation is considered  \citep{plewa,khajenabi15}.

Clouds embedded in a hot gaseous medium have also been discovered near to the Galactic center \citep[][]{Gill}. These low-mass gas clouds, known as G1 and G2, are moving on highly eccentric orbits through gaseous medium around a central black hole.  Using orbital motions of these clouds, one can probe the accretion flow feeding Sgr ${\rm A}^{\star}$ \citep{McCourt2015,Pfuhl2015}. There are considerable uncertainties  about the true nature of intercloud medium and physical mechanisms that may lead to the formation of these clouds. Although various processes have been proposed for the formation of G1 and G2 or BLR clouds, we don't yet know  for sure if these clouds are formed as a result of such mechanisms. The intercloud medium of G1 and G2 is described, however, successfully using a kind of accretion flow which is known as Advection-Dominated Accretion Flow (ADAF; Narayan \& Yi 1994). It is a good motivation to assume that BLR clouds are also moving through this type of accretion flows \citep{krause11}. Nevertheless, most of the previous semi-analytical studies of BLR clouds' dynamics prescribe pressure profile of the intercloud medium as a simple power-law function of the radial distance \citep[e.g.,][]{netzer2010,krause11,krause12,plewa,khajenabi15}. 

Since each BLR cloud is assumed to be in a pressure-confined state, its radius is determined by a balance between the interior pressure of a cloud and its  ambient pressure. Then, orbital motion of a BLR cloud is treated like a classical two-body problem where a cloud with a fixed mass is subject to the central gravity and a force due to the radiation. Most of the previous analytical studies of BLR cloud's dynamics actually follow this approach. 

\cite{netzer2010} studied orbital motion of pressure-confined BLR clouds  in AGNs considering  the combined influence of the central gravity and the radiation pressure. A modified estimate for the mass of the central black hole is presented according to their orbital analysis. \cite{krause11} addressed stability of the orbits using analytical calculations for both isotropic and anisotropic light sources and found that stable orbits may exist under certain circumstances. Although it is unlikely to obtain analytical solutions for the orbital motion of BLR clouds under general conditions, an interesting analytical solution for the orbit of BLR clouds with a fixed column density has been obtained by \cite{plewa}. In all these works, the intercloud is a simple power-law prescription not based on a physically supported model. Moreover, variations of the intercloud's pressure profile with the polar angle has been neglected. These issues motivated \cite{khajenabi15} to study orbital motion of BLR clouds through an ADAF atmosphere where its pressure profile is  based on a two-dimensional self-similar analytical solution for ADAFs  \citep{shadmehri14}. Under these conditions it was shown that stability of the orbits implies that the ensemble of clouds tends to have a disc like configuration.

None of the above studies about orbital motion of BLR clouds considered interaction of the clouds with the surrounding gas via a drag force. As for the G1 and G2 clouds, recent studies show that the drag force has a vital role in the orbits of these clouds \citep[e.g.,][]{McCourt2015,Pfuhl2015}. Just recently, \cite{shadmehri15} studied orbits of BLR clouds subject to a drag force proportional to the velocity. For a particular set of the input parameters, \cite{shadmehri15} presented an analytical solution for the orbits of the clouds which reduces to the analytical solution of \cite{plewa} in the absence of the drag force. In the presence of the drag force, irrespective of the input parameters, orbit of a BLR cloud would decay in a way that it will eventually fall onto the central region. According to the arguments of \cite{shadmehri15} if the time that takes a cloud to reach from its initial position to the central part,  or time-of-flight,  becomes less than   the lifetime of the whole system, then BLR clouds are transient structures rather than long-lived objects so that mechanisms for continually forming BLR clouds are needed. In other words, drag force implies a physical constraint for analyzing orbits of the clouds. \cite{shadmehri15} found that  time-of-flight of a BLR cloud is proportional to the inverse of the dimensionless drag coefficient and using this relation he showed that time-of-flight is indeed shorter than the lifetime of the whole system for a wide range of the input parameters. This interesting finding implies existence of mechanisms for continuously forming these clouds.

However, there are caveats regarding to the analysis of \cite{shadmehri15}. First of all, in his study the drag force is proportional to the velocity which is valid as long as the intercloud is laminar. Although he argues that Reynolds number is less than one which confirms the adopted drag force, for some other input parameters one can easily show that Reynolds number could be  much larger than one. Introducing Reynolds number as ${\rm Re}=\rho v L /\mu$, it can be rewritten as ${\rm Re}=2(v/\bar{u})(L/\lambda)$, where $\rho$, $v$, $L$, $\mu$, $\bar{u}$ and $\lambda$ are the density of gas, the mean velocity of the cloud relative to the gas, characteristic length, dynamic viscosity, the average molecular speed and the mean free path, respectively. If the number density of the intercloud gas is $10^4$ cm$^{-3}$ and its average temperature is $10^8$ K, then we have $\bar{u}\approx 2\times 10^6$ m s$^{-1}$ and $\lambda \approx 10^{10}$ m. The Keplerian velocity at the radial distance 1 pc from the central mass $10^6$ solar mass is around $6.5\times 10^4$ m s$^{-1}$. If we adopt velocity of a cloud approximately equal to this Keplerian velocity and for a typical length $10^{13}$ cm, the Reynolds number becomes around 0.65. Obviously, if the typical length is taken larger, say $10^{15}$ cm, then we have ${\rm Re} = 65$. Also, for a more massive central mass, the Reynolds number is larger than unity. It means that the intercloud medium is turbulent and the drag force should be taken in proportion to the velocity square. In the present work, we plan to study orbits of BLR clouds with a quadratic drag force. At variance with previous work, nevertheless, the intercloud is prescribed using the standard ADAF solutions. Under these circumstances, we calculate time-of-flight of the clouds to see if the main finding of \cite{shadmehri15} is still valid when the drag force is a quadratic function of the velocity. Moreover, in most previous studies, the background gas is assumed to be in a static configuration. We also consider rotation of the medium which a cloud moves through it. In next section, we present basic assumptions and the orbital equations. In section 3, time-of-flight is calculated numerically. We then conclude with our main findings in section 4.

\section{General Formulation}
\subsection{Basic Assumptions}
We study orbital motion of a BLR cloud with mass $m$ subject to three main forces, i.e. gravitational force of a central black hole with mass $M$,  a non-isotropic force due to the radiation of a central accretion disc \citep{Liu11}, and  a drag  force in the opposite direction of the BLR orbital motion. Under these circumstances, direction of cloud's angular momentum is conserved, though its magnitude gradually decreases because of the resistive force. Therefore, motion of a BLR cloud will be in a plane where its inclination $i$ is fixed by the initial angular momentum and it is an input parameter in our model.  A system of coordinates $(x,y,z)$ is constructed so that the central black hole locates at its origin and the central radiating thin accretion is at $z=0$ plane (Figure 1). Thus, location of a cloud in its orbit with inclination angle $i$ with respect to the $x-y$ plane is uniquely determined by its distance $r$ from the origin  and the polar angle $\theta$. The cloud orbit intersects the $x - y$ (disk)plane at the ascending node $A$ so that we define the angle $\angle AOC=\psi$. Having the inclination angle $i$, position of a cloud is determined by $r$ and $\psi$.

\begin{figure}\label{fig:f1}
\includegraphics[scale=0.45]{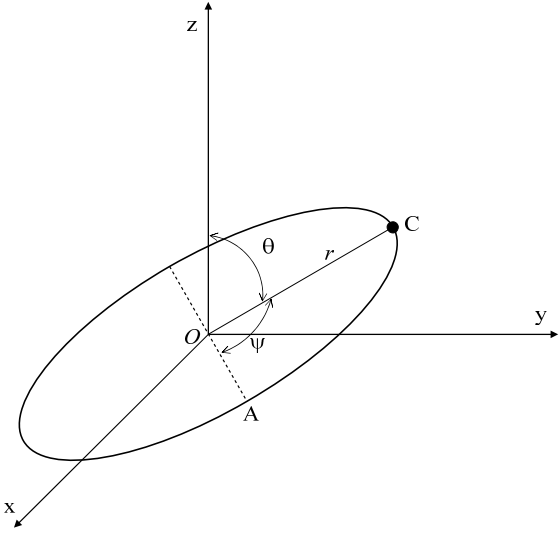}
\caption{The central black hole locates at origin $O$ and position of the cloud C is determined by the radial distance $r$ and the polar angle $\theta$ and the inclination angle of the orbital plane. Since direction of the angular momentum is conserved, the orientation of the orbital plane is fixed and its intersection with the $x-y$ plane is denoted by OA. Thus, position of the cloud C is equivalently determined by the radial distance $r$ and the angle $\angle AOC=\psi$.}
\end{figure}

Geometrical shape of a BLR cloud is assumed to be spherical, for simplicity. Moreover, the clouds are considered to be optically thick. But since the clouds are pressure-confined by definition, physical properties of the ambient gaseous medium like its pressure profile determines how the radius of a cloud varies depending on its position in orbital motion. There are considerable uncertainties about the true nature of intercloud medium.  In other words, irrespective of the confinement mechanisms, internal pressure of a cloud is in balance with the ambient pressure. 

Since  each cloud is in pressure equilibrium with the hot background gas, its radius becomes  $R_{\rm cl}\ \propto   P_{\rm gas}^{-1/3}$ where $P_{\rm gas}$ is
the intercloud gas pressure. On the other hand, according to the standard similarity solutions for
ADAFs \citep{narayan94}, the pressure  distribution is proportional to a power-law function
of the radial distance as $P_{\rm gas}  \propto r^{-s}$ where $s$ is 5/2.
 Therefore, we can rewrite radius of a single  cloud as a function of  its location, i.e.
\begin{equation}
R_{\rm cl} = R_{\rm cl0} \left ( \frac{r}{r_0}\right )^{5/6},
\end{equation}
where $r_0$ is  the initial radial distance of the cloud, and $R_{\rm cl0}$ is the radius  of the cloud at $r_0$. We can calculate the column density $N_{\rm cl} = 3m/2\mu_{\rm m} m_p A $, where $\mu_{\rm m}$ is the mean molecular weight and $A$ is cross section of a cloud.  Having the above relations for  $R_{\rm cl}$ and $P_{\rm gas}$, the column density of a pressure-confined cloud becomes 
 $N_{\rm cl} \propto P_{\rm gas}^{2/3}$ or $N_{\rm cl}=N_0 (r/r_0 )^{-5/3}$
where $N_0 = 3m/(2\mu_{\rm m} m_p \pi R_{\rm cl0}^2)$  is a constant column density. Also,  the density of gas in the standard ADAF similarity solution is written  as a power-law function of the radial distance, i.e.
\begin{equation}\label{eq:adaf-density}
\rho = \rho_0 \left( \frac{r}{r_0} \right)^{-3/2} ,
\end{equation}
where  $\rho_0$ is the  mass density of the intercloud gas at radius $r_0$. We adopt the outer radius of the system as $r_0$ with a value  between 0.01 pc to 1 pc and the number density is $n_0 \sim 10^4 $ cm$^{-3}$ according to the observations \citep[e.g.,][]{Rees89,netzerbook,marconi,plewa}.

The radial velocity $v_{r}$ and the rotational velocity $v_{\varphi}$ of an ADAF are also power-law function of the radial distance. Similarity solutions of \cite{narayan94} are written $v_{r}=-v_{0r} v_{K}$ and $v_{\varphi}=v_{0\varphi} v_{K}$, where $v_{K}=\sqrt{GM/r}$ is Keplerian velocity and the coefficients  $v_{0r}$ and $v_{0\varphi}$ are obtained as,
\begin{equation}\label{eq:vr}
 v_{0r}=(5+2\varepsilon ')\frac{g(\alpha,\varepsilon ')}{3\alpha},
\end{equation}
and
\begin{equation}\label{eq:vphi}
 v_{0\varphi}=\sqrt{\frac{2\varepsilon '(5+2\varepsilon')}{9\alpha^2}g(\alpha,\varepsilon ')} .
\end{equation}
Here, we have 
\begin{equation}\label{eq:ga}
 g(\alpha,\varepsilon ')=\sqrt{1+\frac{18\alpha^{2}}{(5+2\varepsilon ')^2}}-1 ,
\end{equation}
where $\alpha$ is the standard Shakura-Sunyaev viscosity parameter for modeling ADAF's turbulence. Moreover,  the parameter $ \varepsilon '$ is written as $ \varepsilon '=\varepsilon/f$ where
 $\varepsilon = (5/3-\gamma)/(\gamma-1)$, and $\gamma$ is the ratio of specific heats and the parameter $f$ measures the amount of the advected energy.  For example, in a fully advective flow with $\alpha =0.1$ and $\epsilon ' =1$, we obtain $v_{0r} \simeq 0.04$ and $v_{0\varphi} \simeq 0.53$.

\subsection{Equations of Motion in a Static Atmosphere}
We can now write equations of motion of a cloud which is under the influence of the three main forces: the gravitational  force  of the central  mass, ${\bf F}_{\rm grav}$, a force due to non-isotropic radiation of the central accretion disc \citep{Liu11}, ${\bf F}_{\rm rad}$, and a drag force proportional to the velocity square in the opposite direction of the cloud's motion, ${\bf F}_{\rm drag}$. These forces can be written as 
\begin{equation}
{\bf F}_{\rm grav} =- \frac{GMm}{r^2} {\bf e}_{\rm r},
\end{equation}
\begin{equation}
{\bf F}_{\rm rad} = \frac{A}{c} \frac{L_{\rm a}}{2\pi r^2} |\cos \theta | {\bf e}_{\rm r},
\end{equation}
\begin{equation}
{\bf F}_{\rm drag} = -\frac{1}{2} \rho C_D A |{\bf v}| {\bf v},
\end{equation}
where $A$ is the cross sectional area of a cloud and $L_{\rm a}$ is the luminosity of the central source. In the above equation for the drag force, $C_{D}$  is the drag coefficient which  depends on the shape of the cloud and even Reynolds number \citep[e.g.,][]{fluid}. For a sphere, value of $C_D$ may vary from large values for laminar flow to 0.47  for turbulent flow \citep{fluid}.

It is more convenient to re-write the force due to the radiation in terms of the column density $N_{\rm cl}$ and the Eddington ratio $l= L_{\rm a}/ L_{\rm edd}$, where  $L_{\rm edd} = 4\pi GM m_p c /\sigma_{\rm T}$ is  the Eddington luminosity. Here, $\sigma_{\rm T}$ is the Thompson cross-section. Thus, the force due to the radiation becomes 
\begin{equation}
{\bf F}_{\rm rad} = \frac{GMm}{r^2} \frac{3l}{\mu_{\rm m} N_{\rm cl} \sigma_{\rm T}} |\cos\theta | {\bf e}_{\rm r},
\end{equation}
or 
\begin{equation}
{\bf F}_{\rm rad} = \frac{GMm}{r^2} k |\sin \psi | {\bf e}_{\rm r},
\end{equation}
where the dimensionless parameter $k$ is defined as
\begin{equation}
k = \frac{3l}{\mu_{\rm m} N_{\rm cl} \sigma_{\rm T}} \sin (i) .
\end{equation}
Substituting  radial dependence of the column density, the parameter $k$ becomes $k=k_0  (r/ r_0 )^{5/3}$ where $k_0 = 3l \sin (i) /{\mu_{\rm m}} N_0 \sigma_{\rm T}$.

Most of previous authors assume that the intercloud is static which means the background gas does not move. Thus, the relative velocity of a cloud with respect to its ambient medium is cloud's velocity itself. We first consider this simplified situation. Thus, equations of the orbital motion are written as
\begin{equation}\label{eq:main11}
\ddot{r} - r\dot{\psi}^{2} = \frac{GM}{r^2} \left ( k |\sin \psi | - 1 \right ) - \frac{\rho  C_{D}  A}{2}
 \dot{r} \sqrt{\dot {r}^2+r^2 \dot{\psi}^2}  ,
\end{equation}
\begin{equation}\label{eq:main22}
r \ddot{\psi} + 2 \dot{r} \dot{\psi} = - \frac{\rho  C_{D}  A}{2}  r \dot{\psi}  \sqrt{\dot {r}^2+r^2 \dot{\psi}^2}  ,
\end{equation}
where $\dot{r}=dr/dt$, $\ddot{r}=d^2 r/dt^2$ and $\dot{\psi}=d\psi /dt$. Note that  the temperature of a cloud  during its orbital motion is almost constant.

The above orbital equations (\ref{eq:main11}) and (\ref{eq:main22}) are now written in the non-dimensional forms which are more convenient for the numerical integration. Thus, we use the initial radial distance $r_0$ as a reference length scale. Then, Keplerian velocity at this radial distance is written as $v_{\rm K}(r_0 ) = \sqrt{GM/ r_0 }$ and our unit time becomes $t_0 = r_0 / v_{\rm K}(r_0 )$. We now change the variables as $r = r_0  \tilde{r}$ and $t = t_0 \tau $. Thus, equations (\ref{eq:main11}) and (\ref{eq:main22}) become 
\begin{equation}\label{eq:main1}
\ddot{\tilde{r}} -\tilde{r} \dot{\psi}^2 = \frac{1}{\tilde{r}^2}\left(k_0 \tilde{r}^{5/3}|\sin\psi | - 1 \right )-\beta \dot{\tilde{r}} \tilde{r}^{\frac{1}{6}}\sqrt{\dot{\tilde{r}}^2+\tilde{r}^2\dot{\psi}^2},
\end{equation}
and
\begin{equation}\label{eq:main2}
\tilde{r}\ddot{\psi} + 2\dot{\tilde{r}}\dot{\psi} = -\beta \dot{\psi}\tilde{r}^{\frac{7}{6}}\sqrt{\dot{\tilde{r}}^2+\tilde{r}^2\dot{\psi}^2},
\end{equation}
 where $\dot{\tilde{r}}=d\tilde{r}/d\tau $, $\ddot{\tilde{r}}=d^2 \tilde{r}/d\tau^2 $ and $\dot{\psi}=d\psi /d\tau $. The  dimensionless drag coefficient is denoted by $\beta$, i.e.
\begin{equation}\label{eq:beta}
\beta = \frac{3}{8}C_{D}\left ( \frac{r_{0}}{R_{\rm cl0}} \right ) \left (\frac{\rho_0}{\rho_{\rm cl0}} \right ).
\end{equation}

In writing the above equations, we assume that the mass of cloud is conserved during its orbital motion. We note that radius of a cloud and its density at the distance $r_0$ are denoted by $R_{\rm cl0}$ and $\rho_{\rm cl0}$, respectively. Equations (\ref{eq:main1}) and (\ref{eq:main2}) are our main equations for determining orbit of a cloud in a static atmosphere when the drag force is proportional to the velocity square. In section 3, we solve these equations to analyze orbits of a cloud.

\subsection{Equations of Motion in a Rotating Atmosphere}
We now consider a more realistic situation where the ambient gas is rotating and has radial velocity according to equations (\ref{eq:vr}) and (\ref{eq:vphi}). In writing the drag force, then, the relative velocity is considered. Thus, orbital equations become
%
\begin{displaymath}
\ddot{r} - r\dot{\psi}^{2} = \frac{GM}{r^2} \left ( k |\sin \psi | - 1 \right ) - \frac{\rho  C_{D}  A}{2}\times
\end{displaymath}
\begin{equation}\label{eq:main11b}
(\dot{r}+v_{0r}v_{K}) \sqrt{(\dot {r}+v_{0r}v_{K})^2+(r \dot{\psi}-v_{0\varphi}v_{K}})^2,
\end{equation}
\begin{displaymath}
r \ddot{\psi} + 2 \dot{r} \dot{\psi} = - \frac{\rho  C_{D}  A}{2}\times 
\end{displaymath}
%
\begin{equation}\label{eq:main22b}
 (r \dot{\psi}-v_{0\varphi}v_{K} )\sqrt{(\dot {r}+v_{0r}v_{k})^2+(r \dot{\psi}-v_{0\varphi}v_{K}})^2 .
\end{equation}
\begin{figure}\label{fig:f2}
\includegraphics[scale=0.45]{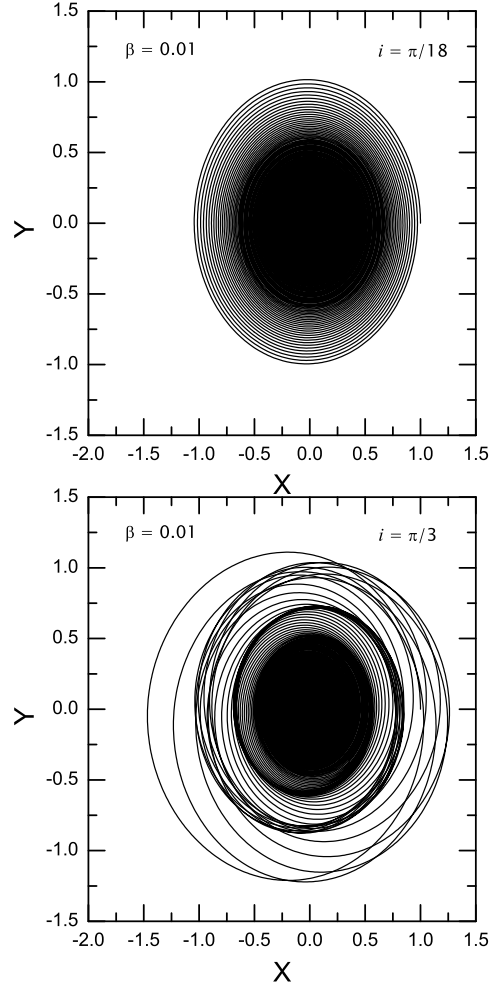}
\caption{Orbit of a BLR cloud in the plane of motion with inclination angles $i=\pi /18$ (top) and $i=\pi /3$ (bottom) for $\beta =0.01$. The other input parameters are $\alpha =0.1$, $\epsilon' =1$, ${\mu_{\rm m}} N_0 \sigma_{\rm T} =3/2$ and $l =0.1$. Initial conditions are $\tilde{r}(\tau =0 )=1$, $\dot{\tilde{r}} (\tau =0)=0$, $\psi (\tau =0) =0$ and $\dot{\psi} (\tau =0)=1$.}
\end{figure}
\begin{figure}\label{fig:f3}
\includegraphics[scale=0.45]{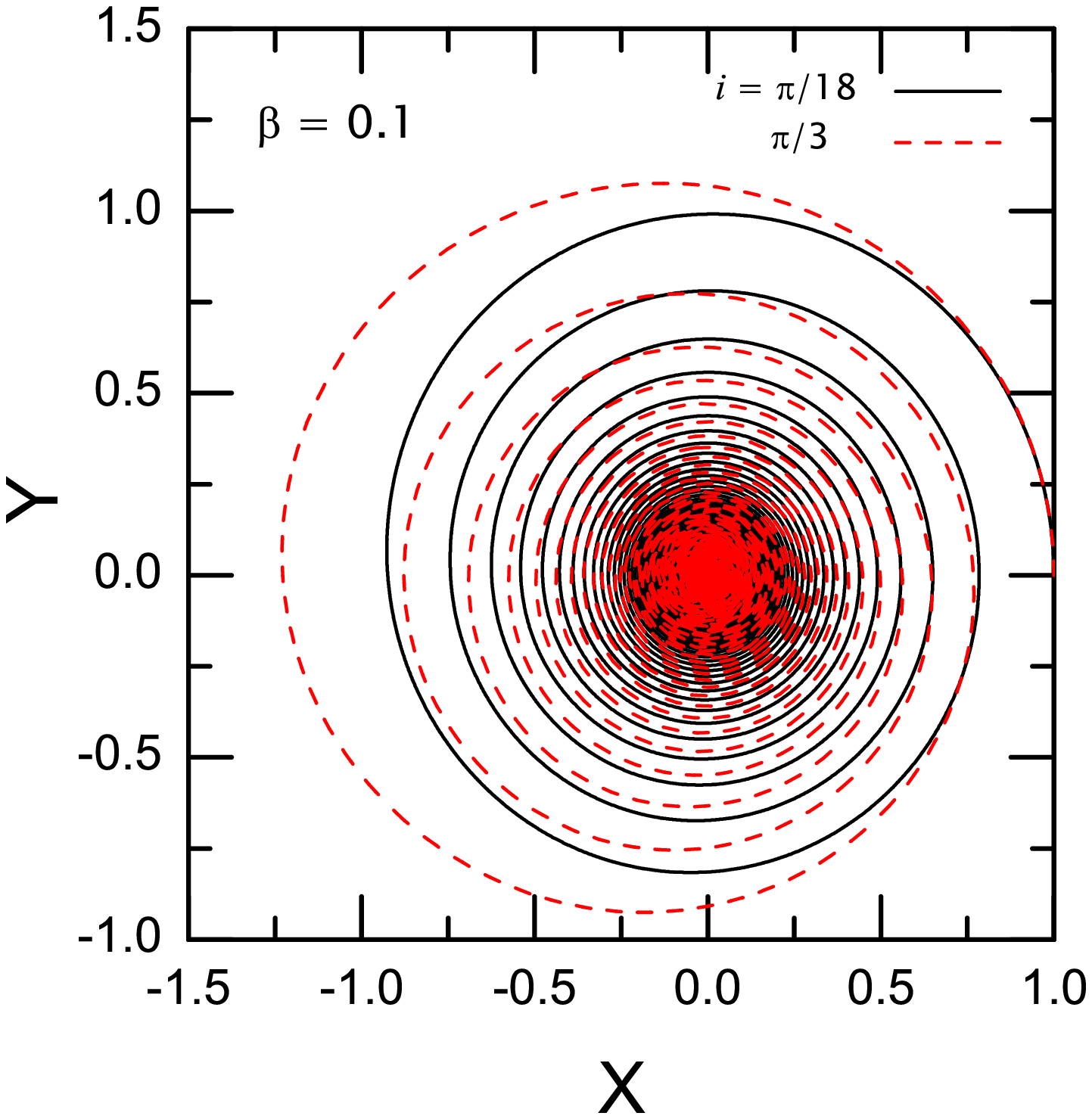}
\caption{Same as Figure 2, but with inclination angles $i=\pi /18$ (solid) and $i=\pi /3$ (dashed) and a larger dimensionless drag coefficient, i.e. $\beta =0.1$. }
\end{figure}
\begin{figure}\label{fig:f4}
\includegraphics[scale=0.45]{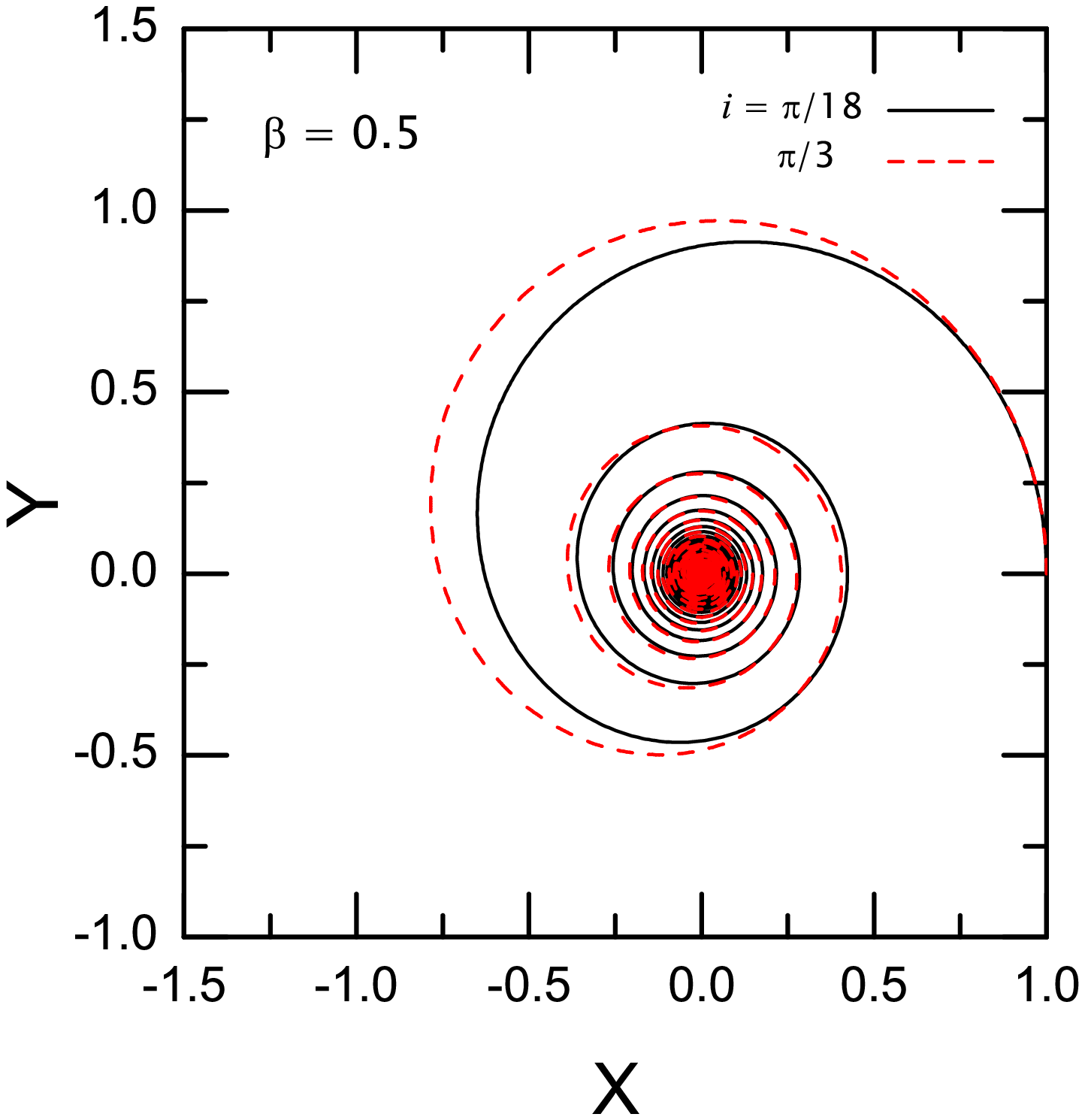}
\caption{Same as Figure 2, but with inclination angles $i=\pi /18$ (solid) and $i=\pi /3$ (dashed) and  a larger dimensionless drag coefficient, i.e. $\beta =0.5$.}
\end{figure}

\begin{figure}\label{fig:f5}
\includegraphics[scale=0.45]{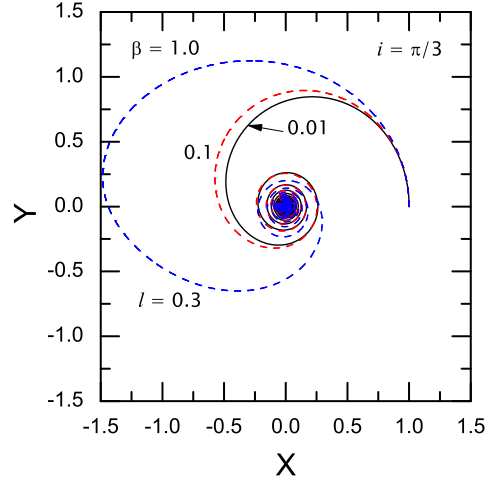}
\caption{ Orbit of a BLR cloud in the plane of motion for cases with different values of the parameter $l$. Here, we have $\beta =1$ and $i=\pi /3$ and the rest of the input parameters and the initial conditions are similar to figure 2.}
\end{figure}

\begin{figure}\label{fig:f6}
\includegraphics[scale=0.45]{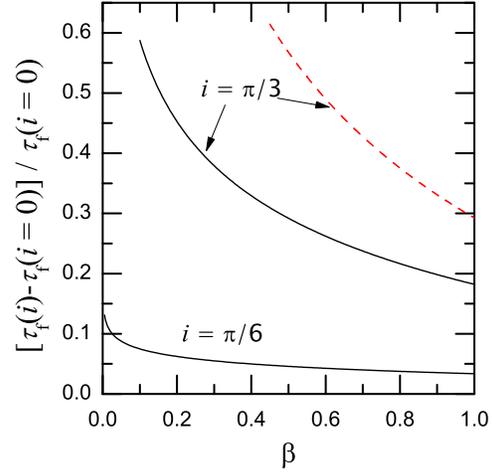}
\caption{ Ratio of $\left [ \tau_f (i) - \tau_f (i=0) \right ] /\tau_f (i=0)$  as a function of the parameter $\beta$ when the intercloud is static (solid) or rotating (dashed) for different inclination angles. Input parameters are $\alpha =0.1$, $\epsilon' =1$, ${\mu_{\rm m}} N_0 \sigma_{\rm T} =3/2$ and $l =0.25$. The initial conditions are $\tilde{r}(\tau =0 )=1$, $\dot{\tilde{r}} (\tau =0)=0$, $\psi (\tau =0) =0$ and $\dot{\psi} (\tau =0)=1$. Each curve is labeled with corresponding value of $i$. When the intercloud is static and the inclination angle is $i=\pi /6$ (solid), there is an increase in the time-of-flight over the case when the inclination angle is zero (i.e., radiation forces do not operate) as the dimensionless drag coefficient decreases.   This enhancement in the time-of-flight is more significant for the smaller values of the dimensionless drag coefficient. Exactly the same behavior is found when the intercloud is rotating and the inclination angle is $i=\pi /6$,  but its difference with the shown case with static atmosphere (solid) is negligible and for this reason, this particular case has not been shown here. }
\end{figure}

 Again,  it is  more convenient to use non-dimensional equations instead of the  above orbital
 equations. So, we transform equations (\ref{eq:main11b}) and (\ref{eq:main22b}) to the following non-dimensional forms:
  \begin{displaymath}
\ddot{\tilde{r}} - \tilde{r}\dot{\psi}^{2} = \frac{1}{\tilde{r} ^2}\left(k_0 \tilde{r}^{5/3}|\sin\psi | - 1 \right) - \beta\times
\end{displaymath}
\begin{equation}\label{eq:main111}
(\dot{\tilde{r}}+v_{0r}\tilde{r}^{-1/2})  \tilde{r}^{\frac{1}{6}}\sqrt{(\dot {\tilde{r}}+v_{0r}\tilde{r}^{-1/2})^2+(\tilde{r} \dot{\psi}-v_{0\varphi}\tilde{r}^{-1/2})^2},
\end{equation}
\begin{displaymath}
\tilde{r} \ddot{\psi} + 2 \dot{\tilde{r}} \dot{\psi} = -\beta\times 
\end{displaymath}
%
\begin{equation}\label{eq:main222}
 ( \dot{\psi}-v_{0\varphi}\tilde{r}^{-3/2}) \tilde{r}^{\frac{7}{6}} \sqrt{(\dot {\tilde{r}}+v_{0r}\tilde{r}^{-1/2})^2+(\dot{\tilde{r}} -v_{0\varphi}\tilde{r}^{-1/2})^2},
\end{equation}
where    dimensionless parameter $\beta$  is defined in equation(\ref{eq:beta}). The above equations are solved subject to the appropriate initial conditions in the next section. 

\begin{table}
\caption{ Our input parameter for describing a BLR cloud and its ambient gas and the resulting dimensionless drag coefficient. Note that each row corresponds to a cloud with a certain mass.}
\begin{tabular}{|l|l|l|l|l| }
\hline
$r_0$ & $n_0$ & $R_{\rm cl0}$ & $n_{\rm cl0}$ & $\beta / C_D $ \\ \hline
1 pc & $10^4$ cm$^{-3}$ & $10^{14}$ cm & $10^{10}$ cm & $1.15\times 10^{-2}$  \\ \hline
1 pc & $10^4$ cm$^{-3}$ & $10^{12}$ cm & $10^{10}$ cm & $1.15$  \\ \hline
0.01 pc & $10^4$ cm$^{-3}$ & $10^{14}$ cm & $10^{10}$ cm & $1.15\times 10^{-4}$  \\ \hline
0.01 pc & $10^4$ cm$^{-3}$ & $10^{12}$ cm & $10^{10}$ cm & $1.15\times 10^{-2}$  \\ \hline
\end{tabular}
\end{table}

\begin{table}
\caption{ Same as Table 1, but each row corresponds to a BLR cloud with a mass equal to $10^{-8}$ solar mass.}
\begin{tabular}{|l|l|l|l|l| }
\hline
$r_0$ & $n_0$ & $R_{\rm cl0}$ & $n_{\rm cl0}$ & $\beta / C_D $ \\ \hline
1 pc & $10^4$ cm$^{-3}$ & $10^{14}$ cm & $4.6\times 10^{6}$ cm & $25$  \\ \hline
1 pc & $10^4$ cm$^{-3}$ & $10^{12}$ cm & $4.6\times 10^{12}$ cm & $2.5\times 10^{-3}$  \\ \hline
0.01 pc & $10^4$ cm$^{-3}$ & $10^{14}$ cm & $10^{10}$ cm & $0.25$  \\ \hline
0.01 pc & $10^4$ cm$^{-3}$ & $10^{12}$ cm & $10^{10}$ cm & $2.5\times 10^{-5}$  \\ \hline
\end{tabular}
\end{table}

\section{ANALYSIS}
We now examine orbits of the BLR clouds in the plane of motion by solving the orbital equations. The background gas is rotating according to ADAF solutions. Describing the results is easier if the same initial conditions are used for all the considered cases. Thus, we assume that a cloud starts its journey  from the initial location $\tilde{r}=1$ (note that all variables are dimensionless). The rest of the initial conditions are  $\dot{\tilde{r}}(\tau =0)=0$, $\psi (\tau =0)=0$ and $\dot{\psi}(\tau =0)=1$. We found that shape of the orbits is qualitatively similar to when a linear drag is used, i.e. orbit of a BLR cloud decays due to the resistive nature of the drag force.  But we can  calculate the time-scale of this orbital decay when the drag force is quadratic. In doing so, time-of-flight $\tau_f$ is defined as the  time needed for traveling of a cloud from its initial location to the center. In order to determine the orbital shape of a BLR cloud and its time-of-flight, we have to adopt the input parameters consistent with the observational data. According to the observations, we have $r_0 \sim 0.01-1 $ pc, $R_{\rm cl0} \sim 10^{14}$ cm, $n_0 \sim 10^4 $ cm$^{-3}$ and $n_{\rm cl0} \sim 10^{10}$ cm$^{-3}$ \citep[e.g.,][]{Rees89,netzerbook,marconi,plewa}. Tables 1 and 2 summarize our input parameters; however, in Table 2 the mass of a BLR cloud is assumed to be  $10^{-8}$ ${\rm M}_{\odot}$.

Figure 2 displays orbital shape of a BLR cloud in the plane of motion (i.e., $XY$-plane where $X$-axis is along OA in Figure 1) with inclination angles $i=\pi /18$ (top) and $i=\pi /3$ (bottom) for dimensionless drag coefficient $\beta =0.01$. Input parameters are $\alpha =0.1$, $\epsilon' =1$ and $k_0 =0.2\sin i$. Also, the initial conditions are $\tilde{r}(\tau =0 )=1$, $\dot{\tilde{r}} (\tau =0)=0$, $\psi (\tau =0) =0$ and $\dot{\psi} (\tau =0)=1$. Radial distance of a cloud gradually decreases because of considering the drag force. The non-isotropic nature of the central radiation becomes more significant with increasing the inclination angle $i$. In Figures 3 and 4 orbital motion of clouds with the same initial and  input parameters are explored but with larger values for the drag coefficient. 

In our model, the effect of the radiation force on the orbit of a cloud appears through the dimensionless parameter $k_0$ which is directly proportional to the Eddington ratio $l$ and $\sin i$. Thus, radiation force operates more significantly in cases with a high inclination angle or a  large Eddington ratio. In order to have bound orbits, the radiation force can not be arbitrary large and for a given set of the input parameters, however, there is always a maximum critical value of parameter $k_0$ so that beyond this value the gravitational force is not able to keep a cloud in a bound orbit.  In Figures  2-4 the orbits are shown for two values of inclination. Since radiation force pushes a cloud toward larger radii, one can expect cases with a larger inclination angle exhibit wider orbits in comparison to a case with a smaller inclination angle. This speculation has been confirmed in Figures 2-4. The effect of the Eddington ratio on the shape of orbits are explored in Figure 5 for different values of the Eddington ratio $l$. Here, we have $\beta =1$ and $i=\pi /3$ and the remaining  input parameters are similar to Figure 2. Having all the parameters fixed, we found that the orbits are no longer bound once the Eddington ratio exceeds a value around 0.3. Nevertheless, the shape of orbits is not modified significantly so long as the ratio $l$ is roughly less than 0.1. Corresponding to the cases with $l=0.01, 0.1$ and 0.3, the dimensionless time-of-flight is found 7.47, 7.78 and 12.75, respectively.

We explored various cases with different sets of the input parameters and the corresponding dimensionless time-of-flight $\tau_f$ is obtained. Interestingly, we found that time-of-flight is in proportion to the inverse of the dimensionless drag coefficient $\beta$ so the constant of the proportionality depends on the input parameters. For a static intercloud gas, we found that $\tau_f (i=0) \approx 2.13 / \beta^{0.89}$, $\tau_f (i=\pi/6) \approx 2.20 / \beta^{0.90}$ and $\tau_f (i=\pi/3) \approx 2.52 / \beta^{1.01}$. When the intercloud is rotating, the time-of-flight is obtained as $\tau_f (i=0) \approx 7.38 / \beta^{0.97}$, $\tau_f (i=\pi/6) \approx 8.00 / \beta^{0.96}$ and $\tau_f (i=\pi/3) \approx 9.54 / \beta^{1.20}$. Except for the cases with zero inclination angle where the radiation force does not  operate, however, in other cases the above fitted time-of-flight functions are not valid for the whole range of the dimensionless drag coefficient $\beta$. For a static intercloud gas, a BLR cloud will not be in bound orbit once the parameter $\beta$ drops to values less than 0.005 for $i=\pi/6$ and 0.1 for $i=\pi/3$. In a rotating intercloud gas, these critical values are larger so that we do not observe bound orbits when $\beta$ is less than 0.02 for $i=\pi/6$ and 0.43 for $i=\pi/3$. Figure 6 shows the ratio $\left [ \tau_f (i) - \tau_f (i=0) \right ] /\tau_f (i=0)$  as a function of the parameter $\beta$ when the intercloud is static (solid) or rotating (dashed) for different inclination angles.

 Thus, we can write $\tau_f \simeq \tau_0 /\beta$ where $\tau_0$ depends on the input parameters.  Although the constant of proportionality $\tau_0$ depends on the input parameters, we found that its variations with the input parameters does not affect significantly the main conclusion in our subsequent discussions.  Time-of-flight for a linear drag is also proportional to inverse of the dimensionless drag coefficient, though definition of this coefficient is different from ours (see Eq.(8) in \cite{shadmehri15}).  One can easily confirm our approximate relation for the time-of-flight using dimensional analysis. A cloud  loses its kinetic energy $1/2 mv^2$ due to the dissipative nature of the drag force with a rate equal to $F_{d}v$, where $v$ is the velocity of the cloud and $m$ is its mass and $F_d$ is the drag force. Thus, time-of-flight can be written as $(1/2 mv^2)/(1/2 \rho C_D  \pi R_{\rm cl}^2 v^3)$ which implies the dimensionless time-of-flight to be proportional to $\beta^{-1}$, i.e. $\tau_f \propto \beta^{-1}$.

For a cloudy BLR system around a black hole with mass $10^8$ M$_{\odot}$, our time unit becomes $t_0 \simeq 1.5$ yr if we set $r_0 = 0.01$ pc. Tables 1 and 2 show that the dimensionless drag coefficient varies from $2.5\times 10^{-5} C_D$ to $25 C_D$ depending on  the background gas density and the properties of a cloud such as its density and radius. Obviously, the longest cloud flight times occur when the parameter $\beta$ is as small as permissible and the radiation force is as large  as it can be. The  effect of the radiation force does not appear for clouds with zero inclination angle and considering the above fitted functions for the time-of-flight, this time-scale will be between $\tau_f (\beta =10^{-5}) \simeq 3\times 10^6$ yr and $\tau_f (\beta =10) \simeq 0.2$ yr for a static intercloud gas. These estimates are modified in a rotating background gas as  $\tau_f (\beta =10^{-5}) \simeq  10^7$ yr and $\tau_f (\beta =10) \simeq 0.7$ yr. For clouds with non-zero inclination angles, however, radiation pressure force increases $\tau_f$ as we confirmed in Figures 5 and 6. But in these cases, there is always a lower limit for $\beta$ so that for the drag coefficient less than this critical value clouds will be pushed outward due to the strong radiation force. When the background gas is static, for example, the explored cases in Figure 6 show that the critical value of $\beta$  is $0.005$ and $0.1$ for inclination angles $\pi /6$ and $\pi /3$, respectively. Then, the time-of-flight becomes $\tau_f (\beta =0.005) \simeq  390$ yr and $\tau_f (\beta =0.1) \simeq  39$ yr which are considerably shorter than the estimated $\tau_f$ for the clouds with zero inclination angle.  Critical value of $\beta$ is larger in a system with a rotating intercloud gas and the corresponding time-of-flight is found as $\tau_f (\beta =0.02, i=\pi/6) \simeq  500$ yr and $\tau_f (\beta =0.43, i=\pi/3) \simeq  40$ yr. 
Using this approximate relation for the time-of-flight, we can discuss about nature of BLR clouds by comparing it with the lifetime of the whole system $\tau_{\rm life}$. If $\tau_{f}$ becomes shorter than $\tau_{\rm life}$, all clouds will fall onto the central object and the system will be depleted of clouds unless  replenishment mechanisms operate to generate new clouds. Observational evidences show that BLRs are clumpy \citep[e.g.,][]{Rees89}, though we do not know if they are continuously forming or long-lived objects. But if $\tau_f < \tau_{\rm life}$, then existence of mechanisms for generating new clouds are needed.  The next step is to obtain a lower limit for the lifetime of the whole system. One can argue so long as a gas reservoir which is known as intercloud gas exists, these BLR clouds may form and move in their orbits. Thus, we can introduce the accretion time-scale as a lower limit for the lifetime of the whole system, i.e. $\tau_{\rm life} = M/\dot{M}$ where $\dot{M}$ is the accretion rate. Despite of uncertainty about the geometry and the nature of the accretion in these system, an approximate relation between the Eddington ratio and the accretion rate can be written as $l\simeq  \dot{M}/\dot{M}_{\rm Edd} $,  where $\dot{M}_{\rm Edd}$ is the Eddington accretion rate \citep[see p. 40,][]{netzerbook}. Thus, one can obtain $\tau_{\rm life} \simeq 4\times 10^8 \frac{\eta}{l}$ yr, where $\eta \sim 0.1$ is the mass-to-luminosity conversion efficiency \citep[][]{netzerbook}. Evidently, this estimated lifetime is much longer than the time-of-flight of BLR clouds except for clouds  with zero inclination angle which may have $\tau_f \sim \tau_{\rm life}$ under very restrictive circumstances. This implies that mechanisms for continuous formation of BLR clouds should operate even when the drag force is quadratic. In the absence of such cloud creation mechanisms, however, it seems only clouds with orbital plane near to equatorial plane may survive and the rest of the clouds will fall onto central black hole very quickly and thereby, a disc like configuration for the geometry of spatial distribution of BLR clouds is expected \citep[also see,][]{khajenabi15}.

Although our model is based on the existence of an ensemble of discrete independent clouds in BLRs, more recent evidence may suggest that the system is not clumpy as has been studied by \cite{arav} in their attempt to find direct signature of discrete clouds in BLR of the Seyfert galaxy NGC 4151. They argued that BLRs are not made of independent clouds. Our analysis shows that even if BLR clouds do exist, they can not be long-lived due to the effect of the drag force. In the absence of a clear physical understanding of possible mechanisms for continuous formation of BLR clouds, however, it seems the system should be depleted of clouds which is consistent with the recent observations \citep{arav}.

 We note that our analysis is based on an assumption which states that the clouds are in pressure equilibrium with their ambient medium. This constraint should not lead to unphysical values for the ratio of density of cloud to the intercloud density, i.e. $\rho_{\rm cl}/\rho$, which  scales as the ratio of inter-cloud medium to cloud temperature. Since mass of cloud $m$ is conserved during its journey, we obtain $\rho_{\rm cl}=\rho_{\rm cl0} (r/r_0 )^{-5/2}$ where $\rho_{\rm cl0} = 3m/(4\pi R_{\rm cl0}^3)\sim 10^{10}$ cm$^{-3}$. Having equation (\ref{eq:adaf-density}) for the intercloud density, we obtain $\rho_{\rm cl}/\rho = \rho_{\rm cl0}/\rho_0 (r/r_0 )^{-1}$ or $\rho_{\rm cl}/\rho \sim 10^6 (r/r_0 )^{-1}$ which means the ratio of the densities can not be arbitrary large so long as a cloud is not very close to the central parts.

\section{Conclusion}
Our goal is to analyze orbits of BLR cloud subject to a drag force proportional to the velocity square. We calculated time-of-flight of the clouds for different initial conditions including motion through a static and rotating atmospheres. In all cases, however, we found that a system is generally older than typical time-scale of spiraling a cloud onto the central region. It means without mechanisms for continuous creation of clouds, a typical BLR system will eventually depleted of clouds. But this feature is not supported by observations. Thus, we can conclude BLR clouds are constantly forming.

\acknowledgments
I am very grateful to referee for his/her constructive report which greatly improved the quality of this paper.

\bibliographystyle{apj}
\bibliography{reference}

\end{document}